\documentclass[12pt]{iopart}

\usepackage{graphicx}
\usepackage{amssymb}

\begin{document}

\letter{Critical Binder cumulant in two--dimensional anisotropic Ising
models}
\author{W Selke$^1$ and L N Shchur$^2$}
\address{$^1$ Institut f\"ur Theoretische Physik, RWTH Aachen, 52056 Aachen, Germany}
\address{$^2$ Landau Institute for Theoretical Physics, 142432 Chernogolovka, Russia}
\ead{selke@physik.rwth-aachen.de}
\submitto{JPA}

\begin{abstract}

The Binder cumulant at the phase transition of Ising models on 
square lattices with various ferromagnetic nearest and
next--nearest neighbour couplings is determined using mainly
Monte Carlo techniques. We discuss the possibility to relate the
value of the critical cumulant in the isotropic, nearest neighbour
and in the anisotropic cases to each other by means of a scale
transformation in rectangular geometry, to pinpoint universal and
nonuniversal features.

\end{abstract}


\section{Introduction}

In the field of phase transitions and critical phenomena,
the fourth order cumulant of the order parameter \cite{Binder}, the
Binder cumulant $U$, plays an important role. Among others, the
cumulant may be used to compute the critical exponent of the
correlation length, and thence to identify the universality class
of the transition.

The value of the Binder cumulant at the transition
temperature, $U(T_c)$, the critical Binder cumulant, has received much
attention, too, indicating the universality
class as well \cite {Privman}. For instance, in the case of the
spin-1/2 Ising model with
ferromagnetic nearest--neighbour couplings the critical cumulant
has been determined very accurately in numerical work, applying
Monte Carlo techniques \cite{Bruce} and transfer-matrix methods
augmented by finite--size extrapolations to the
thermodynamic limit \cite{Bloete}. The resulting
value, $U(T_c)= 0.61069...$\cite{Bloete}, has been observed to hold
also for various other, Ising--type models on a square
lattice \cite{Bruce,Bloete}.

Quite recently, it has been demonstrated by Chen and Dohm \cite{CD} that
the critical Binder cumulant may display non--universal features, when
the couplings on the square lattice are anisotropic. More
specifically, they have studied critical effects of the anisotropy
matrix of the Landau--Ginzburg--Wilson Hamilitonian, consisting
of the coefficients
in front of the leading, second--order gradient term of the
Hamiltonian. A nondiagonal matrix is shown to imply that the
isotropic case cannot be restored by a rescaling of lengths, i. e. in
choosing suitable aspect ratios on a rectangular geometry. Such a
scale transformation is possible in the case of a diagonal anisotropy
matrix, for instance, for nearest--neighbour Ising models with
two different couplings along the axes of the 
square lattice \cite{Bloete,CD,BW}.

In this Letter, we shall present results of rather extensive Monte Carlo
simulations for a variety of anisotropic Ising models on a square
lattice, corresponding to diagonal and nondiagonal anisotropy
matrices. Our data agree qualitatively with the recent results
by Chen and Dohm \cite{CD} on nonuniversality in the case
of a nondiagonal matrix. They confirm quantitatively the scale
transformation given by Kamieniarz and Bl\"ote \cite {Bloete} in
the case of a diagonal matrix.

\section{Results}

We consider spin-1/2 Ising models on a square lattice with
anisotropic nearest and next--nearest neighbour interactions. The 
Hamiltonian reads

 \begin{equation}
{\cal H} = -\sum\limits_{x,y} S_{x,y} ( J_h S_{x+1,y}
 +J_v S_{x,y+1} + J_{d1} S_{x+1,y+1} + J_{d2} S_{x+1,y-1)})
\end{equation}

\noindent
where $S_{x,y}= \pm 1$ is the spin at site $(x,y)$, see Fig. 1. In
most cases, we take
all couplings to be positive, i. e. ferromagnetic. In the Monte Carlo
simulations, systems with $K L$ spins, subject to full periodic
boundary conditions, are analysed, where $K (L)$ corresponds
to the $x (y)$--direction. The aspect ratio, $r$, is defined by 
$r= K/L$.

\begin{figure}
\begin{center}
  \includegraphics[width=0.5\linewidth]{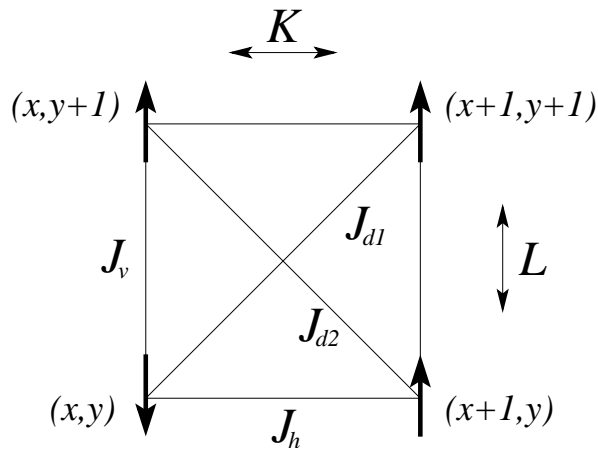}
\end{center}
\label{fig1}
\caption{Sketch of the interactions in the anisotropic Ising model.}
\end{figure}

The critical Binder cumulant is defined by \cite{Binder}

\begin{equation}
  U(T_c) = 1- <M^4>/(3 <M^2>^2)
\end{equation}

\noindent
taking the thermodynamic limit, $L, K \longrightarrow \infty$; 
$<M^2>$ and $<M^4>$ denote the second and fourth moments
of the magnetization, $M= \sum\limits_{x,y} S_{x,y}/(KL)$.

We studied especially two types of anisotropy. In the first
case, where both diagonal next--nearest neighbour couplings
vanish, $J_{d1}= J_{d2}= 0$, the anisotropy matrix is diagonal \cite{CD}.

In the second case, where $J_v= J_h, J_{d1}=J_d$, and 
$J_{d2}= 0$, one encounters nondiagonality \cite{CD}. Indeed, we
consider the two--dimensional
variant of a three--dimensional  model which had been analysed
by Chen and Dohm \cite{CD} and subsequently investigated using Monte
Carlo techniques \cite{Drope}. The simulations, however, may
have been hampered by the fact that the critical
phase transition temperature is not known exactly, in contrast
to the present situation.

Note that data of high accuracy are needed. We
computed systems of sizes, $K, L$, ranging typically from 4 to 64
for square lattices, $K= L$. Using the standard Metropolis
algorithm (a cluster flip algorithm becomes significantly 
more efficient for larger system sizes), Monte Carlo runs with
up to $5\times 10^8$
Monte Carlo steps per site, for the largest systems, were
performed, averaging then over several, about ten, of these
runs to obtain final estimates and to determine the statistical
error bars, shown in the figures. We computed not only the cumulant, but also
other quantities like energy and specific heat, to
check the accuracy of our data. Of course, sufficiently
small lattices may be solved easily by direct enumeration.

\subsection{Diagonal anisotropy matrix}

\begin{figure}
\begin{center}
  \includegraphics[width=0.55\linewidth]{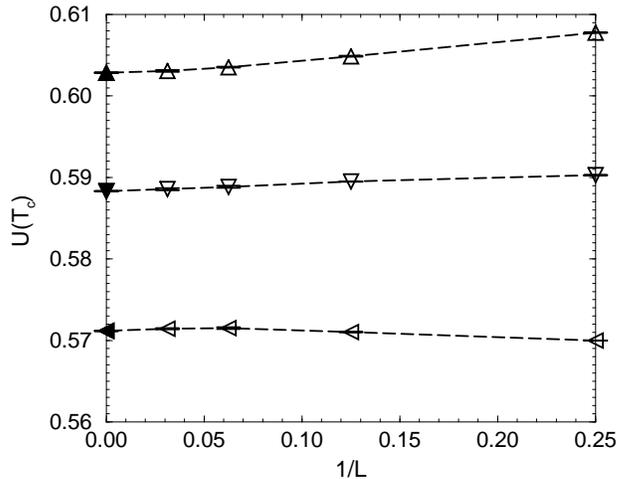}
\end{center}
\label{fig2}
\caption{Critical Binder cumulant $U(T_c)$ for Ising models of size
$L^2$ with $J_v/J_h$= 4.0, 3.0, and 2.0 (from bottom to top). Open
symbols denote Monte Carlo data, full symbols, at $1/L=0$, result
from the scale transformation \cite{Bloete} as described in the
text. Lines are guides to the eye.} 
\end{figure}

We consider the anisotropic Ising model Eq. (1), with
vanishing next--nearest neighbour interactions and
different horizontal, $J_h$, and vertical $J_v$, nearest--neighbour
interactions, see Fig. 1. Then the anisotropy matrix is
diagonal \cite{CD}. The exact transition temperature is given
by \cite{Onsager,Berker} 

\begin{equation}
 \sinh(2J_h/k_BT_c)\sinh(2J_v/k_BT_c)= 1,
\end{equation}

\noindent
where $k_B$ is the Boltzmann constant. Following Kamieniarz
and Bl\"ote \cite{Bloete}, the critical Binder cumulant for
the anisotropic model
on a square lattice, $K= L$, in the thermodynamic limit, can be obtained
from that of the isotropic case, $J_h= J_v$, on a rectangular
lattice with the aspect ratio $r= K/L$ using a 'scale
transformation':

\begin{equation}
 \sinh(2J_h/k_BT_c)= 1/\sinh(2J_v/k_BT_c)=r.
\end{equation}

The critical cumulant $U(T_c)$ of the isotropic case with arbitrary aspect
ratio $r$ has been calculated before, extrapolating to the thermodynamic
limit, with a high degree
of accuracy resulting in a polynomial representation of $U(T_c)$ in
$r$ \cite{Bloete}. Note that this critical cumulant decreases
monotonically from its maximal value at $r=1$  to zero, when the aspect
ratio is increased to infinity \cite{Bloete} (or lowered to zero, because of
the obvious symmetry when replacing $r$ by $1/r$), i. e. when
approaching the one--dimensional limit of the model.  

In Fig. 2 we display results of our check on that scale
transformation, showing our Monte Carlo results of $U(T_c)$ for
anisotropic Ising systems, $J_h \ne J_v$, of small and moderate
sizes, $K=L$. We also include the values we obtained by evaluating
the previously reported polynomial results for the isotropic case with the
appropriate, see Eq. (4), aspect ratio $r$ \cite{Bloete}. We find
a very good, quantitative agreement between our Monte Carlo data
and the values based on the scale transformation and
the polynomial representation \cite{Bloete}.

This observation confirms the universality of the critical
Binder cumulant in the case of a diagonal anisotropy matrix.  

\subsection{Nondiagonal anisotropy matrix}

\begin{figure}
\begin{center}
  \includegraphics[width=0.55\linewidth]{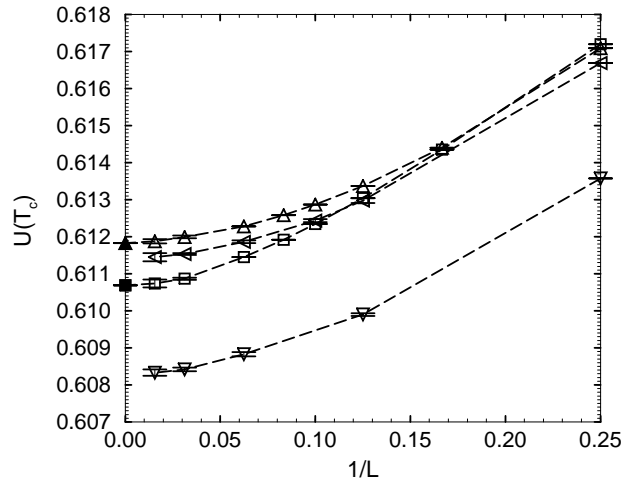}
\end{center}
\label{fig3}
\caption{Critical cumulant of the anisotropic triangular Ising model for
sizes $L^2$, with $J_d/J$= 0.0 (squares), 1.0 (triangles
 up), 1.5 (triangles left), and 2.5 (triangles down). Open symbols denote
our Monte Carlo (as well as numerically exact data for small
lattices), full symbols, at $1/L=0$, denote previous
estimates following from finite--size extrapolations \cite{Bloete}.}
\end{figure}

Now, we consider the case where $J_v= J_h= J$, and $J_{d1}= J_d$,
while $J_{d2}= 0$, see Fig. 1. The model has been called, albeit
being defined on a square lattice, the 'anisotropic triangular
model' \cite{Berker}, because of an obvious isomorphy. Its
anisotropy matrix is nondiagonal \cite{CD}. Again,
the exact transition temperature is known \cite{Berker,Houtappel}

\begin{equation}
 (\sinh(2J/k_BT_c))^2 +2\sinh(2J/k_BT_c)\sinh(2J_d/k_BT_c) = 1,
\end{equation}

\noindent
where $J$ and $J_d$ are supposed to be ferromagnetic in the
following, unless stated otherwise (a weakly
antiferromagnetic $J_d$ does not destroy the ferromagnetic 
order at low temperatures \cite{Berker}). Some of our
simulational results for the Binder cumulant at $T_c$ for square 
lattices, $L=K$, with ferromagnetic couplings are 
shown in Fig. 3, together with highly accurate
estimates based on finite--size extrapolations to the
thermodynamic limit \cite{Bloete} in
the cases $J_d=0$, being the isotropic nearest neighbour square Ising
model, and $J_d=J$, being the isotropic nearest neighbour
triangular Ising model. The Monte Carlo data 
seem to allow a smooth extrapolation to the thermodynamic limit, providing
reliable and accurate estimates (we refrained from a
quantitative finite--size--analysis, because, in general, the form of the
corrections to scaling seems to be unknown).

Results for extrapolations to the thermodynamic limit
are depicted in Fig. 4. Most
importantly, the critical Binder cumulant $U(T_c)$ is seen to increase
from its value in the isotropic, $J_d=0$, case when introducing
the diagonal interaction $J_d > 0$. $U(T_c)$ seems to reach a 
maximum at about $J_d/J$= 1.0, decreasing then, crossing the
isotropic value at about $J_d/J \approx 1.74$ (possibly
at $\sqrt 3$, as one may speculate), and finally
approaching the one--dimensional limit, where $U(T_c)$
tends to go to zero, when $J$ becomes indefinitely weak
compared to $J_d$. As has been just emphasized, $U(T_c)$ is initially, for
$0 < J_d/J \lesssim 1.74$, larger than in the isotropic limit. Now, varying
the aspect ratio and keeping the rectangular geometry
in the isotropic Ising model, this can only lead to
a decrease in the critical Binder
cumulant \cite{Bloete}. Therefore, it
is not possible to obtain the critical Binder cumulant in 
the anisotropic case, $K=L$ and $0 < J_d/J \lesssim 1.74$, by a
scale transformation from the isotropic case, in contrast
to the situation discussed in the
previous subsection. Obviously, this finding is
in accordance with the analysis by Chen and
Dohm \cite{CD}. Our findings do not rule out the
possibility that special cases of the triangular and square
models, choosing suitable
aspect ratios and anisotropies, may be mapped onto each
other, having the same critical cumulants \cite{Bloete}.

Note that the critical cumulant seems to decrease monotonically
when taking a weak antiferromagnetic coupling $J_d$, as we find in
preliminary simulations.

\begin{figure}
\begin{center}
  \includegraphics[width=0.6\linewidth]{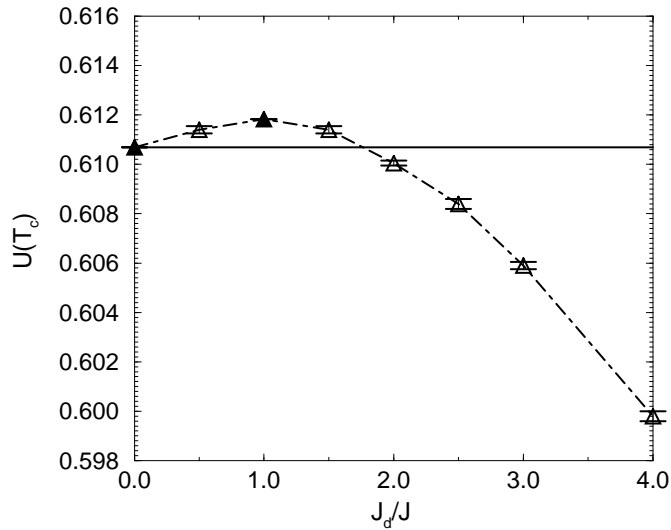}
\end{center}
\label{fig4}
\caption{Critical Binder cumulant, extrapolated to the
 thermodynamic limit, see Fig. 3, for the anisotropic triangular
 Ising model  as function of the anisotropy $J_d/J$. Full symbols
 denote estimates taken from Ref.\cite{Bloete}. The solid line corresponds
 to the isotropic case on a square lattice, $J_d$= 0.}
\end{figure}

Perhaps quite interestingly, the critical Binder cumulant displays
a non--monotonic behaviour when varying the aspect ratio for
the anisotropic triangular model, when $J_d/J$ is sufficiently
large. Starting with the square lattice, $r$=1, and 
then, say, enlargening (or, equivalently due to symmetry, lowering) the
aspect ratio, $U(T_c)$ first becomes larger, before finally
decreasing again as one approaches the one--dimensional geometry.

In summary, the Monte Carlo simulations show that for a diagonal
anisotropy matrix, the critical Binder cumulant in
the anisotropic and isotropic cases are connected by a scale
transformation keeping rectangular geometry. In marked
contrast, such a scale transformation does not exist, in
general, in the anisotropic triangular model described by a nondiagonal
anisotropy matrix, for reasons which have been discussed
recently by Chen and
Dohm, demonstrating non--universal features in
the critical Binder cumulant \cite{CD}. Thence, care
is needed in identifying universality
classes and in estimating phase transition temperatures in using
the critical Binder cumulant. 

\ack
We thank especially V. Dohm for informing us about his work with
X. S. Chen, which, in turn, motivated our work, as well as
for very illuminating discussions. We also thank D. Stauffer for
useful information on the three--dimensional variant of the model 
considered in this Letter.

\end{document}